
\def\bbbc{{\mathchoice {\setbox0=\hbox{$\displaystyle\rm C$}\hbox{\hbox
to0pt{\kern0.4\wd0\vrule height0.9\ht0\hss}\box0}}
{\setbox0=\hbox{$\textstyle\rm C$}\hbox{\hbox
to0pt{\kern0.4\wd0\vrule height0.9\ht0\hss}\box0}}
{\setbox0=\hbox{$\scriptstyle\rm C$}\hbox{\hbox
to0pt{\kern0.4\wd0\vrule height0.9\ht0\hss}\box0}}
{\setbox0=\hbox{$\scriptscriptstyle\rm C$}\hbox{\hbox
to0pt{\kern0.4\wd0\vrule height0.9\ht0\hss}\box0}}}}

\def\bbbr{{\rm I\!R}} 
\font\fivesans=cmss12 at 5pt
\font\sevensans=cmss12 at 7pt
\font\tensans=cmss12
\newfam\sansfam
\textfont\sansfam=\tensans\scriptfont\sansfam=\sevensans
\scriptscriptfont\sansfam=\fivesans
\def\sans{\fam\sansfam\tensans}
\def\bbbz{{\mathchoice {\hbox{$\sans\textstyle Z\kern-0.4em Z$}}
{\hbox{$\sans\textstyle Z\kern-0.4em Z$}}
{\hbox{$\sans\scriptstyle Z\kern-0.3em Z$}}
{\hbox{$\sans\scriptscriptstyle Z\kern-0.2em Z$}}}}

\font \bigbf=cmbx10 scaled \magstep2

\def\slash#1{#1\kern-0.65em /}
\def\dirac{{\raise0.09em\hbox{/}}\kern-0.58em\partial}
\def\Dirac{{\raise0.09em\hbox{/}}\kern-0.69em D}





\magnification=\magstep1
\parskip 4pt plus 1pt
\hoffset=0.1 truecm
\voffset=-0.25 truecm
\hsize=15.5 truecm
\vsize=24.5 truecm
\vglue 1.5cm

\centerline {\bigbf On the Origin of Kaluza-Klein Structure}
\vskip 1.5cm

\centerline {\bf J. Madore}
\medskip
\centerline {\it Laboratoire de Physique Th\'eorique et Hautes
Energies\footnote{*}{\it Laboratoire associ\'e au CNRS.}}
\centerline {\it Universit\'e de Paris-Sud, B\^at. 211,  \ F-91405 ORSAY}
\vskip 1cm

\centerline {\bf J. Mourad}
\medskip
\centerline {\it  Laboratoire de Mod\`eles de Physique Math\'ematique}
\centerline {\it Parc de Grandmont, Universit\'e de Tours, \ F-37200 TOURS}
\vskip 2cm

\noindent
{\bf Abstract:} \ It is suggested that quantum fluctuations of the light
cone are at the origin of what appears at low energy to be a
higher-dimensional structure over space-time. A model is presented which
has but a finite number of Yang-Mills fields although the supplementary
algebraic structure is of infinite dimension.

\vfill
\noindent
LPTHE Orsay 95/00

\noindent
hep-th/9506041
\medskip
\noindent
March, 1995
\bigskip
\eject

\beginsection 1 Introduction

The fact that Yang-Mills potentials can be unified with the
gravitational potential to build the components of a metric in a
higher-dimensional manifold can be taken as evidence that in some sense
the effective dimension of space-time is in fact larger than four. A
modification of the traditional realization of this argument has been
proposed in which the additional dimensions are replaced by an algebraic
structure and an associated differential calculus which is not
necessarily derived from the traditional calculus of a differential
manifold. That is, it has been proposed that the extra structure should
be described by a noncommutative geometry.  (Dubois-Violette {\it et
al.} 1989, Madore 1990, Chamseddine {\it et al.} 1993, Madore \& Mourad
1993).  This means that the Kaluza-Klein unification can be achieved
using a space-time of dimension four. There remains however the problem
of explaining the extra algebraic structure. We suggest here that its
origin is to be found in the quantum fluctuations of the gravitational
field and that it is essentially a quantum effect. Specifically, we show
that in spite of the fact that the quantum fluctuations would be
expected to give rise to an infinite-dimensional algebra and hence {\it
a priori} to an infinite number of Yang-Mills potentials, a model can be
proposed which has only a finite number.

We are not in a position to give a space-time description of the effect
we wish to describe and so we shall restrict our considerations to an
arbitrarily chosen causal space-like slice $V$.  Since our interest is
only in the local structure of space-time, it is not a restriction to
suppose that such a $V$ exists. Once we have argued that $V$ acquires an
extra noncommutative structure we must suppose that time evolution will
induce a similar structure on all of space-time.

We shall see that the restriction on the dimension of the space of
possible Yang-Mills potentials stems from the fact that our definition
of a linear connection makes essential use of the bimodule structure of
the space of 1-forms. This accounts also for the difference of some of
our conclusions from those of authors (Chamseddine {\it et al.} 1993,
Sitarz 1994, Klim\v c\' ik {\it et al.} 1994, Landi {\it et al.} 1994)
who define a linear connection using the classical (Koszul 1960) formula
for a covariant derivative on an arbitrary left (or right) module. A
more detailed comparison of the two approaches is given in Sitarz
(1995).  We refer to Bailin \& Love (1987) for an introduction to
standard Kaluza-Klein theory and, for example, to Madore \& Mourad
(1993) for a motivation of the generalization to noncommutative
geometry.  By a connection we shall always mean here a metric-compatible
torsion-free linear connection.

In Section~2 we give a short review of the algebraic version of
Kaluza-Klein theory and its possible relation to quantum fluctuations.
The material here is mainly speculative. It is partly an adaptation of
the standard folklore (Deser 1957, Isham {\it et al.} 1971) on the
role of the gravitational field as a universal regulator.  In Section~3
we recall the definition of an extension of a linear connection within
the context of noncommutative geometry. This is in part a recapitulation
of the work of Kehagias {\it et al.} (1995). In Section~4 we present two
models with an infinite-dimensional algebra as extra structure but with
only a finite number of extra Yang-Mills modes.

\beginsection 2 Quantum fluctuations

Let $V$ be a causal space-like slice of space-time and let ${\cal C}(V)$
be the (commutative, associative) algebra of smooth functions on $V$.  A
smooth, classical, scalar field defines by restriction an element of
${\cal C}(V)$.  A quantum scalar field $f(x)$ is an operator-valued
distribution on space-time. Although by restriction it does not define a
distribution on $V$, if one smears it over an arbitrarily small
time-like interval it defines a smooth function on $V$ with values in
the unbounded operators ${\cal L}$ on some Hilbert space (Borchers
1964).  Formally it defines by restriction an element of an algebra
${\cal A}_0$ of functions on $V$ with values in ${\cal L}$.  If $x$ and
$y$ are two distinct points of $V$ they have space-like separation with
respect to the Minkowski metric and $f(x)$ and $f(y)$ commute:
$$
f(x) f(y) = f(y) f(x).                                            \eqno(2.1)
$$
We can formally identify ${\cal A}_0$ with the tensor product of
${\cal C}(V)$ and a commutative subalgebra ${\cal L}_0$ of ${\cal L}$:
$$
{\cal A}_0 = {\cal C}(V) \otimes_\bbbc {\cal L}_0.
$$

Suppose now that $V$ is considered as fixed but that the space-time
metric is allowed to vary. If the metric is such that two points $x$ and
$y$ of $V$ have no longer a space-like separation then (2.1) will no
longer necessarily be valid. The commutation relations of a quantum
scalar field depend critically on the space-time metric in which the
field is quantized.  Suppose in particular that because of quantum
effects the metric fluctuates around the Minkowski metric. Let
$\langle f(x) \rangle_g$ be the mean value of $f(x)$ taken over these
fluctuations. Then it is to be expected that
$$
\langle f(x) f(y) \rangle_g \neq \langle f(x) f(y) \rangle_g       \eqno(2.2)
$$
even though $x$ and $y$ have a space-like separation with respect to the
original Minkowski metric. We introduce a new product $*$ on ${\cal L}$
defined by
$$
f_1(x) * f_2(y) = \langle f_1(x) \, f_2(y) \rangle_g               \eqno(2.3)
$$
and we suppose that it is regulated by the gravitational fluctuations in
the sense that the limit
$$
f_1 * f_2 (x) = \lim_{y\rightarrow x} f_1(x) * f_2(y)              \eqno(2.4)
$$
is well defined. In general it is to be expected that
$$
f_1 * f_2 \neq f_2 * f_1.                                          \eqno(2.5)
$$

We define the (noncommutative, associative) algebra ${\cal A}$ to be
the algebra of functions ${\cal A}_0$ but with the product (2.4). We shall
in the following drop the $*$ in the product and we shall consider
${\cal A}$ to be the structure algebra of a classical noncommutative
geometry.  Let $\tilde Z^0$ be the center of ${\cal A}$.  We shall
assume that in some quasi-classical approximation we can identify
$\tilde Z^0$ with the original algebra of functions:
$$
\tilde Z^0 = {\cal C}(V).                                          \eqno(2.6)
$$
The only other general property one could reasonably expect to know of
${\cal A}$ is that the commutator of two of its elements vanishes with
some fundamental length which tends to zero with
$\hbar$(Planck mass)$^{-1}$. We shall however not explicitly use this fact.

{}From (2.6) it follows that there exists a differential algebra (Connes
1986) $\tilde \Omega^* = \Omega^*({\cal A})$ over ${\cal A}$ which
admits an imbedding
$$
\Omega^*(V)
\buildrel i \over \longrightarrow \tilde \Omega^*                  \eqno(2.7)
$$
where $(\Omega^*(V), d)$ is the standard differential calculus over
space-time. When necessary we shall distinguish the differential on
space-time with a subscript $V$. Such an imbedding has been proposed
(Dubois-Violette {\it et al.} 1989, Madore 1990, Madore \& Mourad 1993,
Kehagias {\it et al.} 1995, Madore 1995) as an appropriate
noncommutative generalization of Kaluza-Klein theory. From (2.7) it
follows that the 1-forms $\tilde \Omega^1$ can be written as a direct
sum
$$
\tilde \Omega^1 = \tilde \Omega^1_h \oplus \tilde \Omega^1_c       \eqno(2.8)
$$
where, in the traditional language of Kaluza-Klein theory, $\Omega^1_h$
is the horizontal component of the 1-forms. It can be defined as the
$\tilde \Omega^0$-module generated by the image of $\Omega^1(V)$ in
$\tilde \Omega^1$ under the imbedding; it is given by
$$
\tilde \Omega^1_h =
\Omega^1(V) \otimes_\bbbc \tilde \Omega^0.                         \eqno(2.9)
$$
The $\tilde \Omega^1_c$ is a complement of $\tilde \Omega^1_h$ in
$\tilde \Omega^1$. A vector-space complement always exists. If we
suppose that the quantum fluctuation which give rise to the $*$-product
are about Minkowski space then there is an action of the Poincar\'e
algebra on $\tilde \Omega^*$ and we can write the algebra
$\tilde \Omega^0$ as a product
$$
\tilde \Omega^0 = {\cal C}(V) \otimes_\bbbc \Omega^0               \eqno(2.10)
$$
where $\Omega^*$ is some differential algebra. In this case we can
choose
$$
\tilde \Omega^1_c = \tilde \Omega^0 \otimes_\bbbc \Omega^1         \eqno(2.11)
$$
and $\tilde \Omega^1_c$ is an $\tilde \Omega^0$-module complement of
$\tilde \Omega^1_h$. The horizontal component of the 1-forms can be
expected to have a more general significance whereas the complement
depends on the Ansatz (2.10). The differential algebra $\Omega^*$
replaces the hidden manifold of traditional Kaluza-Klein theory.

Our starting point in the next section will be the Formula~(2.6) which
can be considered to be to a certain extent independent of the
speculations which preceded it.  Let $\tilde Z^1$ be the vector space of
elements of $\tilde \Omega^1_c$ which commute with $\tilde \Omega^0$.
The most important assumption we make is that the dimension of $\tilde
Z^1$ is finite.  This assumption is completely unmotivated. It is
introduced to account for the fact that there are a finite number of
observed Yang-Mills potentials. It replaces the truncation assumptions
which are introduced in traditional Kaluza-Klein theories but it is a
weaker assumption since in general the space $\tilde \Omega^1_c$ itself
will be of infinite dimension as a vector space and in general not even
finitely generated as a $\tilde \Omega^0$-module.

\beginsection 3 Covariant derivatives

Consider the case of a general (algebraic) tensor product
$$
\tilde \Omega^* = \Omega^{\prime *} \otimes_\bbbc \Omega^{\prime\prime *}
                                                                   \eqno(3.1)
$$
of two arbitrary differential calculi $\Omega^{\prime *}$ and
$\Omega^{\prime\prime *}$ with corresponding covariant derivatives
$D^\prime$ and $D^{\prime\prime}$. We have then two linear maps
$$
\Omega^{\prime 1} \buildrel D^\prime \over \longrightarrow
\Omega^{\prime 1} \otimes_{\Omega^{\prime 0}} \Omega^{\prime 1},    \qquad
\Omega^{\prime\prime 1} \buildrel D^{\prime\prime} \over \longrightarrow
\Omega^{\prime\prime 1} \otimes_{\Omega^{\prime\prime 0}}
\Omega^{\prime\prime 1},                                           \eqno(3.2)
$$
which satisfy the corresponding Leibniz rules. It has been stressed
(Dubois-Violette \& Michor 1995, Mourad 1995) that the appropriate
generalization to noncommutative geometry of the Leibniz rule involves a
generalized symmetry operation $\sigma$. We have then two maps
$$
\Omega^{\prime 1} \otimes_{\Omega^{\prime 0}}
\Omega^{\prime 1} \buildrel \sigma^\prime \over \longrightarrow
\Omega^{\prime 1} \otimes_{\Omega^{\prime\prime 0}}
\Omega^{\prime 1},                                               \qquad
\Omega^{\prime\prime 1} \otimes_{\Omega^{\prime\prime 0}}
\Omega^{\prime\prime 1} \buildrel \sigma^{\prime\prime}
                                   \over \longrightarrow
\Omega^{\prime\prime 1} \otimes_{\Omega^{\prime\prime 0}}
\Omega^{\prime\prime 1}
$$
which are necessarily bilinear. In simple models, the quantum plane
(Dubois-Violette {\it et al.} 1995) and some matrix models (Madore {\it
et al.} 1995), the covariant derivative and the associated map $\sigma$
have been shown to be essentially unique.

{}From $D^\prime$ and $D^{\prime\prime}$ we wish to construct an extension
$$
\tilde \Omega^1 \buildrel \tilde D \over \longrightarrow
\tilde \Omega^1 \otimes_{\tilde \Omega^0} \tilde \Omega^1,         \eqno(3.3)
$$
with its associated generalized symmetry operation $\tilde \sigma$.
This is the most general possible formulation of the bosonic part of the
Kaluza-Klein construction under the Ansatz (3.1). It has been shown
(Kehagias {\it et al.} 1995) that the existence of a non-trivial
extension places severe restrictions on the differential calculi of the
two factors. These restrictions are at the origin of the fact that we
can construct a Kaluza-Klein theory using an infinite algebra but still
with only a finite number of Yang-Mills modes.

The 1-forms $\tilde \Omega^1$ can be written as in (2.8) with, in the
present notation, (2.9) and (2.11) given by
$$
\tilde \Omega^1_h =
\Omega^{\prime 1} \otimes_\bbbc \Omega^{\prime\prime 0}, \qquad
\tilde \Omega^1_c =
\Omega^{\prime 0} \otimes_\bbbc \Omega^{\prime\prime 1}.
$$
Let $f^\prime \in \Omega^{\prime 0}$,
$f^{\prime\prime} \in \Omega^{\prime\prime 0}$ and
$\xi^{\prime} \in \Omega^{\prime 1}$,
$\xi^{\prime\prime} \in \Omega^{\prime\prime 1}$.  Then it follows
from the definition of the product in the tensor product that
$$
f^\prime \xi^{\prime\prime} = \xi^{\prime\prime} f^\prime, \qquad
f^{\prime\prime} \xi^\prime = \xi^\prime f^{\prime\prime}.         \eqno(3.4)
$$
Hence one concludes that the extension $\tilde\sigma$ of $\sigma^\prime$
and $\sigma^{\prime\prime}$ which is part of the definition of
$\tilde D$ is given by
$$
\tilde\sigma(\xi^\prime \otimes \eta^{\prime\prime}) =
\eta^{\prime\prime} \otimes \xi^\prime,                 \qquad
\tilde\sigma(\xi^{\prime\prime} \otimes \eta^\prime) =
\eta^\prime \otimes \xi^{\prime\prime}.                            \eqno(3.5)
$$
{}From these one deduces the constraints (Kehagias {\it et al.} 1995)
$$
f^\prime \tilde D \xi^{\prime\prime} =
(\tilde D \xi^{\prime\prime}) f^\prime,\qquad
f^{\prime\prime} \tilde D \xi^\prime =
(\tilde D \xi^\prime) f^{\prime\prime}                             \eqno(3.6)
$$
on $\tilde D$. These are trivially satisfied if
$$
\tilde D\xi^\prime = D^\prime \xi^\prime,                \qquad
\tilde D\xi^{\prime\prime} = D^{\prime\prime} \xi^{\prime\prime}.  \eqno(3.7)
$$

Using the decomposition (2.8) one sees that the covariant derivative
(3.3) takes its values in the sum of 4 spaces, which can be written in
the form
$$
\eqalign{
&\tilde \Omega^1_h \otimes_{\tilde \Omega^0} \tilde \Omega^1_h =
(\Omega^{\prime 1} \otimes_{\Omega^{\prime 0}} \Omega^{\prime 1})
\otimes_\bbbc \Omega^{\prime\prime 0},                          \cr
&\tilde \Omega^1_h \otimes_{\tilde \Omega^0} \tilde \Omega^1_v =
\Omega^{\prime 1} \otimes_\bbbc \Omega^{\prime\prime 1},        \cr
&\tilde \Omega^1_v \otimes_{\tilde \Omega^0} \tilde \Omega^1_h =
\Omega^{\prime\prime 1}  \otimes_\bbbc \Omega^{\prime 1},       \cr
&\tilde \Omega^1_v \otimes_{\tilde \Omega^0} \tilde \Omega^1_v =
\Omega^{\prime 0} \otimes_\bbbc (\Omega^{\prime\prime 1}
\otimes_{\Omega^{\prime\prime 0}} \Omega^{\prime\prime 1}).     \cr
}                                                                  \eqno(3.8)
$$
Let $Z^{\prime 0}$ ($Z^{\prime\prime 0}$) be the center of
$\Omega^{\prime 0}$ ($\Omega^{\prime\prime 0}$) and let  $Z^{\prime 1}$
($Z^{\prime\prime 1}$) be the vector space of elements of
$\Omega^{\prime 1}$ ($\Omega^{\prime\prime 1}$) which commute with
$\Omega^{\prime 0}$ ($\Omega^{\prime\prime 0}$). Then $Z^{\prime 1}$
($Z^{\prime\prime 1}$) is a bimodule over $Z^{\prime 0}$
($Z^{\prime\prime 0}$). Let $Z^{\prime 2}$ ($Z^{\prime\prime 2}$)
be the elements of
$\Omega^{\prime 1} \otimes_{\Omega^{\prime 0}} \Omega^{\prime 1}$
($\Omega^{\prime\prime 1} \otimes_{\Omega^{\prime\prime 0}}
\Omega^{\prime\prime 1}$) which commute with $\Omega^{\prime 0}$
($\Omega^{\prime\prime 0}$). Then one finds the inclusion relations
$$
Z^{\prime 1} \otimes_{Z^{\prime 0}} Z^{\prime 1}
\subset Z^{\prime 2},                                     \qquad
Z^{\prime\prime 1} \otimes_{Z^{\prime\prime 0}} Z^{\prime\prime 1}
\subset Z^{\prime\prime 2},
$$
but in general the two sides are not equal.  From (3.6) we see then
that
$$
\eqalign{
&\tilde D\xi^\prime \in
(\Omega^{\prime 1} \otimes_{\Omega^{\prime 0}} \Omega^{\prime 1})
\otimes_\bbbc Z^{\prime\prime 0}                          \oplus
\Omega^{\prime 1} \otimes_\bbbc Z^{\prime\prime 1}        \oplus
Z^{\prime\prime 1} \otimes_\bbbc \Omega^{\prime 1}        \oplus
\Omega^{\prime 0} \otimes_\bbbc Z^{\prime\prime 2},       \cr
&\tilde D\xi^{\prime\prime} \in
Z^{\prime 0} \otimes_\bbbc (\Omega^{\prime\prime 1}
\otimes_{\Omega^{\prime\prime 0}} \Omega^{\prime\prime 1})\oplus
\Omega^{\prime\prime 1} \otimes_\bbbc Z^{\prime 1}        \oplus
Z^{\prime 1} \otimes_\bbbc \Omega^{\prime\prime 1}        \oplus
Z^{\prime 2} \otimes_\bbbc \Omega^{\prime\prime 0}.
}                                                                  \eqno(3.9)
$$
In the relevant special case with
$\Omega^{\prime *} = \Omega^*(V)$, we shall have
$$
Z^{\prime 0} = \Omega^{\prime 0},  \qquad
Z^{\prime 1} = \Omega^{\prime 1},                                  \eqno(3.10)
$$
and so (3.9) places no restriction on $\tilde D\xi^{\prime\prime}$.

The main assumption which we shall make is that
$$
{\dim}_\bbbc (Z^{\prime\prime 1}) < \infty.                        \eqno(3.11)
$$
We shall also suppose that
$$
Z^{\prime\prime 1} \otimes_{Z^{\prime\prime 0}} Z^{\prime\prime 1}
= Z^{\prime\prime 2}                                               \eqno(3.12)
$$
although it is easy to find pertinent cases where this would not be so.
For example, the matrix geometry introduced by Connes \& Lott (1992) has
a vanishing left-hand side and a right-hand side of dimension 1. We
refer to Kehagias {\it et al.} (1995) for details. Let $\theta^a$ be a
basis of $Z^{\prime\prime 1}$ over the complex numbers.  From (3.12) we
can conclude that
$$
d\theta^a = -{1\over 2} C^a{}_{bc} \theta^b \theta^c.              \eqno(3.13)
$$
The product on the right-hand side is the product in the algebra
$\Omega^{\prime\prime *}$. The $C^a{}_{bc}$ are elements of the algebra
$\Omega^{\prime\prime 0}$. Since the left-hand side commutes with all
elements of the algebra they lie in the center $Z^{\prime\prime 0}$
of $\Omega^{\prime\prime 0}$. We shall suppose that they are complex
numbers.

We shall impose the condition that the connections be metric and without
torsion although these might be considered rather artificial conditions
on the components of the 1-forms in $\tilde \Omega^1_c$. We have then
two bilinear maps
$$
\Omega^{\prime 1} \otimes_{\Omega^{\prime 0}} \Omega^{\prime 1}
\buildrel g^\prime \over \longrightarrow \Omega^{\prime 0}, \qquad
\Omega^{\prime\prime 1} \otimes_{\Omega^{\prime\prime 0}}
\Omega^{\prime\prime 1}
\buildrel g^{\prime\prime} \over \longrightarrow
\Omega^{\prime\prime 0},                                           \eqno(3.14)
$$
which satisfy the compatibility condition (1.9), from which we must
construct an extension
$$
\tilde \Omega^1 \otimes_{\tilde \Omega^0} \tilde \Omega^1
\buildrel \tilde g \over \longrightarrow \tilde \Omega^0
$$
which satisfies also (1.9). From the decomposition (3.8) one sees that
$\tilde g$ will be determined by two bilinear maps
$$
\Omega^{\prime 1} \otimes_\bbbc \Omega^{\prime\prime 1}
\buildrel g_1 \over \longrightarrow \tilde \Omega^0,  \qquad
\Omega^{\prime\prime 1} \otimes_\bbbc \Omega^{\prime 1}
\buildrel g_2 \over \longrightarrow \tilde \Omega^0.
                                                                   \eqno(3.15)
$$
If $\tilde g$ is symmetric then from (3.5) it follows that
$$
g_2 = g_1 \tilde \sigma.
$$
In general it is to be expected that if the connection is metric and
without torsion, the conditions (3.11) will place constraints also on the
covariant derivative $\tilde D\xi^{\prime\prime}$.

In the relevant special case with $\Omega^{\prime *} = \Omega^*(V)$ one
can define a metric $i^* \tilde g$ on $V$ by
$$
i^* \tilde g(\theta^\alpha, \theta^\beta)
  = \tilde g(\tilde \theta^\alpha, \tilde \theta^\beta),       \quad
    \tilde \theta^\alpha = i(\theta^\alpha).
$$
To maintain contact with the commutative construction of the previous
section we suppose in this case that
$$
i^* \tilde g = g_V.                                                \eqno(3.16)
$$

Let $\theta^\alpha$ be a (local) moving frame on $V$ and set
$\theta^i = (\theta^\alpha, \theta^a)$.  The Kaluza-Klein construction
follows as in Section~3 of Kehagias {\it et al.} (1995).

\beginsection 4 Models

There remains the task of constructing a reasonable model of a
differential calculus which satisfies the conditions (3.11) and (3.12).
Consider the case where the classical theory is invariant under an
internal symmetry group $G$. Then the quantum theory can be defined to be
also invariant under $G$. In particular, there is an action of $G$
on the algebra ${\cal A}$. In order to be able to make use of previous
calculations we shall consider only the case $G = SU_n$. Let $\lambda_a$
be a basis of the Lie algebra of $G$ and $e_a$ the associated
derivations of ${\cal A}$.  There exists then (Dubois-Violette 1988) a
differential calculus $(\tilde \Omega^*_D, \tilde d_D)$ based on the
derivations such that $\tilde \Omega^1_D$ has a basis $\theta^a$
(Dubois-Violette {\it et al.} 1989, 1990) with the property that it
commutes with the elements of ${\cal A}$. That is $\tilde \Omega^*_D$ is
of the form (3.1) with
$$
Z^{\prime\prime 1}_D = \{ \theta^a \}.                               \eqno(4.1)
$$
Therefore (3.11) and (3.12) are satisfied.  Let now
$(\tilde \Omega^*, \tilde d)$ be an arbitrary differential calculus
which is an extension of $(\tilde \Omega^*_D, \tilde d_D)$ and which is
such that
$$
Z^{\prime\prime 1} = Z^{\prime\prime 1}_D, \qquad
Z^{\prime\prime 2} = Z^{\prime\prime 2}_D.                           \eqno(4.2)
$$
Then (3.11) and (3.12) remain satisfied for
$(\tilde \Omega^*, \tilde d)$. The extended differential calculus would
have to be `large enough', be such that the condition $\tilde d f = 0$,
for $f \in {\cal A}$ implies that $f = 1$. This would not be the case
for $(\tilde \Omega^*_D, \tilde d_D)$.

As a second example let us consider the equilibrium physics of a
particle, confined to a two-dimensional plane, which obeys anyon
statistics. The space-time is of euclidean signature then and the
quantum fluctuations of the light-cone are replaced by the equally
complicated effects of solid-state physics. The classical wave function
can be considered as a function on $\bbbr^2$ with values in the quantum
plane. We shall formally identify the algebra ${\cal A}$ of such
functions with the tensor product of the algebra of functions on
$\bbbr^2$ and the algebra of the quantum plane. As differential calculus
$\Omega^*({\cal A})$ we can choose the tensor product of the form (3.1)
with $\Omega^{\prime *} = \Omega^*(\bbbr^2)$ and
$\Omega^{\prime\prime *}$ equal to the differential calculus on the
quantum plane introduced by Pusz \& Woronowicz (1989) and Wess \&
Zumino (1990).  We have then a Kaluza-Klein theory over $\bbbr^2$.
Following the reasoning of Kehagias {\it et al.} (1995) one can show
that there is only a trival Kaluza-Klein extension. This is not the case
if one replaces the quantum plane with the noncommutative torus (Connes
\& Rieffel 1987, Connes 1994). The algebra ${\cal A}_{\alpha}$ of the
latter is an infinite involutive unital algebra generated by two
elements $u$ and $v$ subject to the relations:
$$
vu=e^{2\pi i\alpha}uv,\qquad u^* = u^{-1}, \quad v^* = v^{-1},      \eqno(4.3)
$$
where $\alpha$ is an irrational number. The
differential calculus based on the Dirac operator (Connes \& Rieffel
1987) is equivalent to the differential calculus based on the outer
derivations. These are the derivations modulo the inner derivations.
They have two generators
$$
e_{1}(u^{n}v^{m})=inu^{n}v^{m},\qquad
e_{2}(u^{n}v^{m})=imu^{n}v^{m}.                                     \eqno(4.4)
$$

The space $\Omega^{\prime\prime 1}$ is a free ${\cal A}_{\alpha}$-bimodule
with a basis $\theta^{a}$ dual to $e_{a}$.  An arbitrary 1-form
may be written as $\omega = \omega_{a}\theta^{a}$ with $\omega_{a}$
two elements of ${\cal A}_{\alpha}$.  It is easily verified that
$$
Z^{\prime\prime 1}=\{\theta^{a}\},\qquad
Z^{\prime\prime 2}=\{\theta^{a}\otimes\theta^{b}\},                 \eqno(4.5)
$$
and that
$$
d\theta^{a}=0.                                                      \eqno(4.6)
$$
The modified second example is similar to the first one but with the
group $G$ equal to $U_1\times U_1$.

\beginsection 5 Conclusions

We have presented a rather general formulation of Kaluza-Klein theory in
which the internal manifold is replaced by an abstract differential
calculus with the unique restriction that the dimension of the space of
1-forms which commute with all elements of the algebra be finite. This
restriction replaces the truncation assumption in traditional
Kaluza-Klein theory. We have shown that with this assumption the number
of Yang-Mills potentials is also finite in spite of the fact that the
internal-structure algebra is otherwise quite arbitrary.

\parskip 7pt plus 1pt
\parindent=0cm
{\it Acknowledgment:}\ The first author would like to thank
H. Goenner and the Heraeus Foundation for financial support during
a three-month visit to the Institut f\"ur Theoretische Physik,
Universit\"at G\"ottingen where part of this research was done. He would
like to thank A. Dimakis and F. M\"uller-Hoissen for some very
lively (lebhafte) discussions there. He would like to thank also M.
Dubois-Violette for enlightening discussions on algebraic field theory.

\vskip 1cm
\vfill\eject

\beginsection References

Bailin D., Love A. 1987, {\it Kaluza-Klein theories}, Rep. Prog. Phys.
{\bf 50} 1087.

Borchers H.J. 1964, {\it Field Operators as ${\cal C}^\infty$ Functions
in Spacelike Directions}, Il Nuovo Cimento {\bf 33} 1600.

Chamseddine A.H., Felder G., Fr\"ohlich J. 1993, {\it Gravity in
Non-Commutative Geometry}, Commun. Math. Phys. {\bf 155} 205.

Connes A. 1986, {\it Non-Commutative Differential Geometry}, Publications
of the Inst. des Hautes Etudes Scientifique. {\bf 62} 257.

--- 1994, {\it Non-commutative geometry}, Academic Press.

Connes A., Lott J. 1990, {\it Particle Models and Noncommutative Geometry},
in `Recent Advances in Field Theory', Nucl. Phys. Proc. Suppl. {\bf B18} 29.

--- 1992, {\it The metric aspect of non-commutative geometry},
Proceedings of the 1991 Carg\`ese Summer School, Plenum Press.

Connes A., Rieffel M. 1987 {\it Yang-Mills for non-commutative two tori}
Contemp. Math. {\bf 62} 237.

Deser S. 1957, {\it General Relativity and the Divergence Problem in
Quantum Field Theory}, Rev. Mod. Phys. {\bf 29} 417.

Dubois-Violette M. 1988, {\it D\'erivations et calcul diff\'erentiel
non-commutatif}, C. R. Acad. Sci. Paris {\bf 307} S\'erie I 403.

Dubois-Violette M., Michor P. 1994, {\it D\'erivations et calcul
diff\'erentiel non-commuta\-tif II},
C. R. Acad. Sci. Paris {\bf 319} S\'erie I 927.

--- 1995, {\it Connections on Central Bimodules}, Preprint LPTHE Orsay 94/100.

Dubois-Violette M., Kerner R., Madore J. 1989 {\it Classical bosons in a
noncommutative geometry}, Class. Quant. Grav. {\bf 6} 1709.

--- 1990, {\it Non-Commutative Differential Geometry of Matrix Algebras},
J. Math. Phys. {\bf 31} 316.

Dubois-Violette M., Madore J., Masson T., Mourad J. 1994, {\it Linear
Connections on the Quantum Plane}, Lett. Math. Phys. {\bf 0} 00.

Isham C.J., Salam A., Strathdee J. 1971, {\it Infinity Suppression in
Gravity-Modified Quantum Electrodynamics}, Phys. Rev. {\bf D3} 1805.

Kehagias A., Madore J., Mourad J., Zoupanos G. 1995, {\it Linear
Connections on Extended Space-Time}, Tours Preprint, LMPM 95-01.

Klim\v c\'ik C., Pompo\v s A., Sou\v cek. V. 1994, {\it Grading of
Spinor Bundles and Gravitating Matter in Noncommutative Geometry},
Lett. Math. Phys.  {\bf 30}, 259.

Koszul J.L. 1960, {\it Lectures on Fibre Bundles and Differential Geometry},
Tata Institute of Fundamental Research, Bombay.

Landi G., Nguyen Ai Viet, Wali K.C. 1994, {\it Gravity and
electromagnetism in noncommutative geometry}, Phys. Lett. {\bf B326} 45.

Madore J. 1990, {\it Modification of Kaluza-Klein Theory},
Phys. Rev. {\bf D41} 3709.

--- 1995, {\it An Introduction to Noncommutative Differential Geometry
and its Physical Applications}, Cambridge University Press.

Madore J., Masson T., Mourad J. 1994, {\it Linear Connections on Matrix
Geometries}, Class. Quant. Grav. {\bf } (to appear).

Madore J., Mourad J. 1993, {\it Algebraic-Kaluza-Klein Cosmology},
Class. Quant. Grav. {\bf 10} 2157.

Mourad. J. 1995, {\it Linear Connections in Non-Commutative Geometry},
Class. Quant. Grav. {\bf 12} 965.

Pusz W., Woronowicz S.L. 1989, {\it Twisted Second Quantization},
Rep. on Math. Phys. {\bf 27} 231.

Sitarz A. 1994, {\it Gravity from Noncommutative Geometry},
Class. Quant. Grav. {\bf 11} 2127.

--- 1995, {\it On some aspects of linear connections in
noncommutative geometry}, Jagiellonian University Preprint.

Wess J., Zumino B. 1990, {\it Covariant Differential Calculus on the Quantum
Hyperplane} Nucl. Phys. B (Proc. Suppl.) {\bf 18B} 302.

\bye